\documentclass[pre,twocolumn,superscriptaddress]{revtex4}
 
\usepackage{epsfig}
\usepackage{verbatim}
\usepackage{bm}
\usepackage{multirow}
\usepackage[colorlinks]{hyperref}
\usepackage{anysize}
\usepackage{relsize}
\usepackage{xcolor}
 \usepackage{newpxtext,newpxmath}

\hypersetup{colorlinks=false,
	linkcolor=red,
	anchorcolor=black,
	citecolor=blue,
	urlcolor=black
}

\marginsize{1.5cm}{1.5cm}{1.cm}{1.cm}

\providecommand{\be}{\begin{equation}}
\providecommand{\ee}{\end{equation}}
\providecommand{\bea}{\begin{eqnarray}}
\providecommand{\eea}{\end{eqnarray}}
\providecommand{\beas}{\begin{eqnarray*}}
\providecommand{\eeas}{\end{eqnarray*}}
\providecommand{\bal}{\begin{aligned}}
\providecommand{\eal}{\end{aligned}}

\providecommand{\bsi}{{\bm \sigma}}

\providecommand{\bS}{{\bf S}}

\providecommand{\bD}{{\bm \Delta}}
\newcommand{\nn}{\nonumber}

\begin{document}

\title[]{Real-space renormalization-group methods for hierarchical spin glasses}
\author{Michele Castellana}
\affiliation{Laboratoire Physico-Chimie Curie, Institut Curie, PSL Research University, CNRS UMR 168, Paris, France.\\
Sorbonne Universit\'es, UPMC Univ. Paris 06, Paris, France }

\begin{abstract}
We focus on two real-space renormalization-group (RG) methods recently proposed for a hierarchical model of a spin glass: A sample-by-sample method, in which the RG transformation is performed separately on each disorder sample, and an ensemble RG (ERG) method [M. C. Angelini, G. Parisi, and F. Ricci-Tersenghi. Ensemble renormalization group for disordered systems. \textit{Phys. Rev. B}, 87(13):134201, 2013] in which the transformation is based on an average over  samples. 
Above the upper critical dimension, the sample-by-sample method yields the correct mean-field value for the critical exponent $\nu$ related to the divergence of the correlation length, while it does not predict the correct qualitative behavior of $\nu$  below the upper critical dimension. On the other hand,  the ERG procedure has been claimed to predict the correct behavior of $\nu$ both above and below the upper critical dimension. 
Here, we straighten out the reasons for the discrepancy between the two  methods above, by demonstrating that the ERG method predicts a marginally stable critical fixed point, thus implying a prediction for the critical exponent $\nu$ given by $2^{1/\nu} = 1$. This prediction disagrees, on a qualitative and quantitative level, both with the mean-field value of $\nu$  above the critical dimension, and with numerical estimates of $\nu$ below the upper critical dimension. Therefore, our results show that finding a  real-space RG method for spin glasses which yields the correct prediction for universal quantities below the upper critical dimension is still an open problem, for which our analysis may provide some general guidance for future studies. 
\end{abstract}

\maketitle

\setlength{\parskip}{5pt plus 0pt minus 0pt}

\section{Introduction}
\label{intro}
The renormalization-group (RG)  has proven to be a ubiquitous, powerful method to study and reduce complex physical systems to a handful of degrees of freedom, which constitute the only few relevant variables in the critical regime \cite{wilson1975renormalization}, and  allowed for predicting the critical features  of a large variety  of  systems, including  binary fluids, superfluids, polymers, and ferromagnetic materials \cite{justin1996quantum}. 

Among the systems in solid-state physics that still lack a full theoretical understanding are spin glasses---disordered uniaxial magnetic materials, such as a  solution of Mn in Cu, modeled by an array of spins on the Mn arranged at random in the Cu matrix \cite{anderson1995through}. Because of their physical and mathematical richness and complexity, spin glasses have been interesting theoreticians for decades now, and  a full physical understanding of their critical and low-temperature features has been obtained only in the mean-field approximation \cite{parisi1980order}.

This mean-field solution involves a functional order parameter characterized by the existence of many pure states \cite{parisi1983order,mezard1984replica}, and it has recently been proven to be correct on a rigorous level \cite{guerra2003broken,talagrand2006parisi}. However, the validity of the mean-field picture above for non-mean-field spin glasses, e.g., finite-dimensional models with short-range interactions, is still debated. In particular, several studies proposed a scenario for the low-temperature phase of non-mean-field spin glasses which markedly differs from the mean-field picture above. In this scenario, the low-temperature phase is characterized by a single ergodic component and by its spin-reversed counterpart \cite{fisher1986ordered,fisher1987absence}. 

Among the causes of the difficulties in solving non-mean-field models of spin glasses is the fact that RG methods  have proven to be challenging when applied to these systems \cite{parisi2001renormalization}, from both the conceptual and technical standpoint. For instance, the RG equations based on the replica approach \cite{mezard1987spin} imply a remarkably complex perturbative structure and, to the best of our knowledge, the  resulting  $\epsilon$-expansion around the upper critical dimension has not proven to be predictive \cite{chen1977mean,castellana2011renormalization}. 

A natural strategy to overcome the issue above is to develop an RG method which is not based on replicas, but on a real-space picture, in which at each step of the RG transformation the degrees of freedom on the system's lattice are first decimated, and then rescaled so as to obtain a new lattice with the same number of degrees of freedom as the original one. 
In this regard, real-space RG methods have been recently applied  to the hierarchical Edwards-Anderson model (HEA) \cite{franz2009overlap}, a spin-glass model built on a hierarchical lattice, whose recursive structure is ideally suited to the real-space RG transformation \cite{dyson1969existence}. 
In this study, we focus on two  real-space RG approaches recently proposed for the HEA: A sample-by-sample procedure in which the decimation  is performed individually for each disordered sample \cite{castellana2011real}, and one in which the decimation involves an average over disordered samples---the ensemble RG (ERG) method \cite{angelini2013ensemble}. The sample-by-sample procedure predicts the correct mean-field value of the critical exponent $\nu$ related to the divergence of the correlation length above the upper critical dimension, while its predictions disagree with numerical simulations below the upper critical dimension \cite{franz2009overlap,angelini2013ensemble}. On the other hand, the ERG procedure has been claimed to predict the correct behavior of $\nu$, both above and below the upper critical dimension \cite{angelini2013ensemble,decelle2014ensemble}. 

In this analysis, we straighten out the reasons for the discrepancy between the two RG procedures above: By examining the fixed-point structure of the ERG method, we demonstrate that the ERG procedure predicts a marginally stable critical fixed point, and that the  largest eigenvalue of the linearized RG transformation at such fixed point is equal to one. By using the relation between such eigenvalue and  $\nu$, we obtain that  the ERG method predicts that $\nu$ is given by the relation $2^{1/\nu} = 1$. This prediction disagrees with both the mean-field value of $\nu$  above the upper critical dimension, and with numerical estimates of $\nu$ below the upper critical dimension \cite{franz2009overlap,angelini2013ensemble}. As a result, our analysis indicates that finding a suitable real-space RG method for spin glasses which yields the correct predictions for universal quantities below the upper critical dimension is still an open problem. 

\section{Results}

 In order to analyze the ERG method, in what follows we introduce the HEA  \cite{franz2009overlap}. The HEA  is a system of Ising spins $S_i = \pm 1$, whose Hamiltonian is defined by the  recursion relation
\be\label{eq_H}
H_{k+1}[\bS]  = H_k[\bS_{\rm L}] + H_k[\bS_{\rm R}] - 2^{-\varsigma (k+1)} \sum_{i<j=1}^{2^{k+1}} J_{ij} S_i S_j,
\ee
where in what follows vector quantities will be written in bold, and $\bS_{\rm L} = \{ S_1, \cdots, S_{2^k}\}$, $\bS_{\rm R} = \{ S_{2^k+1}, \cdots, S_{2^{k+1}}\}$ denote the spins in the left and right half of the system, respectively. The pairwise bonds  $J_{ij}$ at the $k+1$-th hierarchical level are independent,  normally distributed random variables with zero mean and standard deviation $\sigma_{k+1}$, where $\sigma_{k+1}$ will be specified in the RG analysis below. Bonds on different levels are mutually independent, and the initial condition of the recursion relation is $H_0[\bS] = 0$. In the definition (\ref{eq_H}), $H_k[\bS_{\rm L}]$ and $H_k[\bS_{\rm R}]$ denote the Hamiltonians of two subsystems with size $2^k$, which are coupled with an interaction energy given by the third term in the right-hand side (RHS). The exponent $\varsigma$ sets the interaction range: the larger $\varsigma$, the faster the   interaction between a pair of spins decays with respect to their hierarchical distance \cite{franz2009overlap}.  In what follows we will focus on the region $1/2 < \varsigma < 1$, see  \cite{franz2009overlap} and \cite{castellana2011real} for details. 
In addition, we recall that the HEA with a given $\varsigma$ can be approximately mapped into a short-range spin glass on a $d$-dimensional lattice, according to a  correspondence between the exponent $\varsigma$ and the dimension $d$ \cite{banos2012correspondence,angelini2013ensemble}. In particular, the regions $1/2 <\varsigma  < 2/3$ and $2/3 <\varsigma  < 1$ qualitatively correspond to a  dimension $d$ larger and smaller than the upper critical dimension, respectively \cite{franz2009overlap}. 

In the ERG  approach proposed in \cite{angelini2013ensemble}, a HEA with $k$ hierarchical levels and couplings with standard deviations 
\be
\bsi = \{ \sigma_1, \cdots, \sigma_k\}
\ee 
is approximated by a HEA with $k-1$ levels and couplings with standard deviations 
\be
\bsi' = \{ \sigma'_1, \cdots, \sigma'_{k-1}\},
\ee
by imposing a set of equalities on the sample averages of some  observables: 
\be\label{rg}
\mathbb{E}[ O_{l+1} ] = \mathbb{E}[O'_{l} ], \; 1 \leq l \leq k-1,
\ee
where we denote by $\mathbb{E}[]$  the average with respect to  disorder samples. For the  model with $k$ levels,  $\langle \rangle$ is the Boltzmann average at inverse temperature $\beta = 1/T$, 
\be
\langle \cdot \rangle \equiv \frac{1}{Z} \sum_\bS e^{-\beta H_k[\bS]} \; \cdot,
\ee
$Z$ is the  partition function, and $O_l$ the correlation  between a  left and a right block of spins  ${\rm L}_l = \{ 1, \cdots, 2^{l-1} \}$ and ${\rm R}_l = \{ 2^{l-1}+1, \cdots , 2^l \}$ at level $l-1$: 
\beas
O_l  = \frac{\sum_{i \in {\rm L}_l, j \in {\rm R}_l}\langle S_i S_j \rangle^2}{\sqrt{\sum_{i \in {\rm L}_l, j \in {\rm L}_l}\langle S_i S_j \rangle^2\sum_{i \in {\rm R}_l, j \in {\rm R}_l}\langle S_i S_j \rangle^2}}.
\eeas
Proceeding along the same lines, in what follows all quantities related to the model with $k-1$ levels, e.g., $O'_l$ and $\langle \rangle'$,  will be denoted by a $'$, where the Boltzmann average $\langle \rangle '$ is at the inverse temperature $\beta$ defined above.

Given that the left- and right-hand side of Eq. (\ref{rg}) depend on $\bsi$ and $\bsi'$, respectively, Eq. (\ref{rg}) yields a mapping 
\be\label{map}
\bsi \rightarrow \bsi'
\ee
which depends on the inverse temperature $\beta$.  The transformation (\ref{map}) can then be iterated multiple times: at the $t$-th step, we set
\be\label{eq10}
\bsi = \bsi^t,
\ee 
and a $k-1$-level model with standard deviations $\bsi'$ is obtained from  (\ref{map}). Two copies of this decimated model are then coupled according to Eq. (\ref{eq_H}), and a new $k$-level model with standard deviations $\bsi^{t+1}$ is built. In this model,  the couplings at the  $k$th level  are chosen to have the same   standard deviation $\sigma^t_k$ as the original model, Eq. (\ref{eq10}): 
\be\label{rec}
\bsi^{t+1} = \{ \sigma'_1, \cdots, \sigma'_{k-1}, \sigma^t_k \},
\ee
where in the RHS $\sigma'_1, \cdots, \sigma'_{k-1}$ depend on $\bsi^t$, cf. Eqs. (\ref{map}) and (\ref{eq10}). 

\begin{figure*}
\begin{center}
\includegraphics[scale=1]{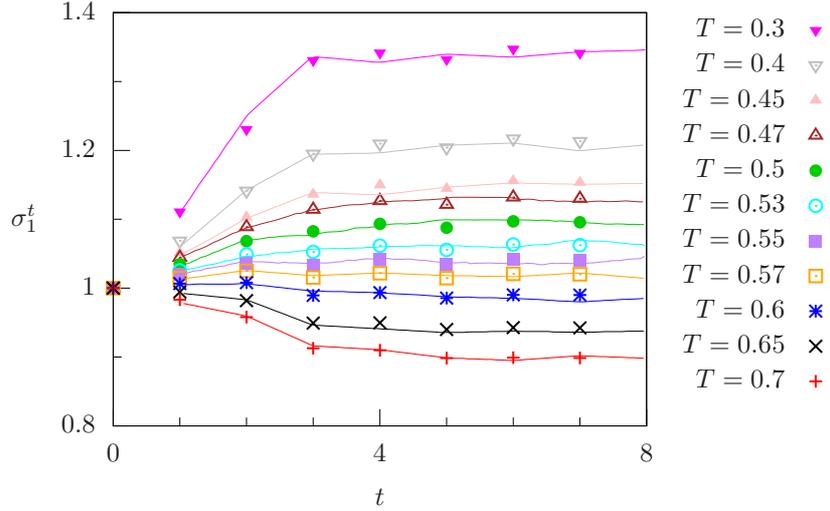}
\caption{\label{fixed_beta}
Iteration of the ensemble renormalization-group  (RG) transformation with $k=4$ hierarchical levels at fixed temperature, for  $\varsigma = 5/6$. Standard deviation $\sigma^t_1$ of the couplings at the first hierarchical level in the 4-level model, as a function of the number $t$ of RG steps for different temperatures. Each temperature corresponds to a color, points have been obtained with our analysis, and lines are from  \cite{angelini2013ensemble}. 
}
\end{center}
\end{figure*}

The RG transformation above can be iterated by solving numerically Eqs. (\ref{rg}) at fixed temperature. In this analysis, we have done this by using stochastic-approximation methods, which yield the solution $\bsi'$ to any degree of accuracy \cite{robbins1951stochastic,blum1954multidimensional}. We choose an initial value for $\bsi'$, draw randomly a disorder sample, correct $\bsi'$ according to an update rule based  on the values  of the observables $O_l$, $O'_{l-1}$ computed on the disorder sample above, and iterate this procedure multiple times, until $\bsi'$ converges to the solution. The results for this iteration obtained with  initial condition 
\be\label{ic}
\sigma^0_i = 1, \; 1 \leq i \leq k,
\ee
and  $\varsigma = 5/6$, which is claimed to approximately correspond to a short-range model in three dimensions \cite{angelini2013ensemble},  are shown Fig. \ref{fixed_beta}, and they reproduce those of \cite{angelini2013ensemble}.

\begin{figure*}
\begin{center}
\includegraphics[scale=.8]{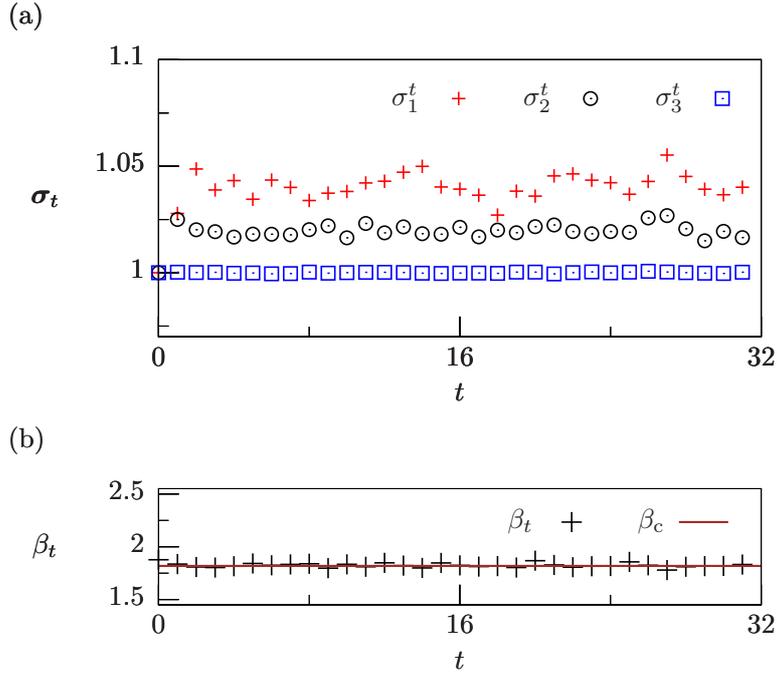}
\caption{\label{beta_cr}
Iteration of the ensemble renormalization-group  (RG) transformation with $k=4$ hierarchical levels and 
$\varsigma = 5/6$, where at each step we solve for the inverse temperature so as to find the critical fixed point. (a) Standard deviations $\sigma^t_1, \cdots, \sigma^t_3$ of the couplings at the three lowest hierarchical levels in the 4-level model (red crosses, black circles and blue squares, respectively) as functions of the number $t$ of RG steps. 
(b) Inverse critical temperature $\beta_t$ as a function of the number of RG steps (black crosses). 
After a few iterations, $\bsi^t$ and $\beta_{t}$ reach the critical fixed point $\bsi_{{\rm c} }$ and inverse critical temperature $\beta_{\rm c}$ (brown line), respectively.}
\end{center}
\end{figure*}

Building on the analysis above, we now determine the critical fixed point and the critical exponent $\nu$. We start with the same initial condition (\ref{ic}) as above and iterate the RG transformation (\ref{rec}): in addition to solving Eq. (\ref{rg}) for $\bsi'$, at each step $t$ we solve for the inverse temperature $\beta$ by requiring that $\sigma'_{k-1} = \sigma^t_k$, i.e., that the standard deviation of the coupling at the last level of the decimated model equals that of the original model. Proceeding along the lines of the iteration at fixed temperature, we solve for $\beta$  with stochastic-approximation methods, and denote the solution by $\beta_t$. The result of this procedure for $\varsigma = 5/6$ is shown in Fig. \ref{beta_cr}: after a few iterations, both the standard deviations  on all levels, $\bsi^t$, and the inverse temperature $\beta_t$ plateau out and reach a critical fixed point $\bsi_{\rm c}$, $\beta_{\rm c}$, respectively. In particular, the critical temperature $T_{\rm c} = 1/\beta_{\rm c} = 0.55(3)$ is consistent with the value reported in  \cite{angelini2013ensemble}.

Finally, we compute the critical exponent $\nu$ by linearizing the RG transformation (\ref{rec}) at the critical fixed point above. By deriving both sides of Eq. (\ref{rg}) with respect to $\sigma_j$, we have
\be\label{lin}
\frac{\partial \mathbb{E}[ O_{l+1} ]}{\partial \sigma_j} = \sum_{i=1}^{k-1} \frac{\partial \mathbb{E}[O'_{l} ]}{\partial \sigma'_i} \frac{\partial \sigma'_i}{\partial \sigma_j}.
\ee
The derivatives of the expectation values can be written explicitly and readily evaluated numerically: for example
\be\label{der}
\frac{\partial \mathbb{E}[ O_{l+1} ]}{\partial \sigma_k} = \frac{1}{\sigma_k} \mathbb{E} \mathlarger{\Bigg[}\mathlarger{\Bigg(} \frac{\sum_{i<j=1}^{2^k} J_{ij}^2}{\sigma_k^2} - \frac{2^k(2^k-1)}{2} \mathlarger{\Bigg)} O_{l+1}\mathlarger{\Bigg]},
\ee
where the sum in the RHS involves only couplings at the $k$th hierarchical level, and similarly for the other derivatives. The matrix relative to the linearized RG equations is 
\bea\label{M}\nn
M_{ij} &=& \frac{\partial \sigma^{t+1}_i}{\partial \sigma^t_j} \\
&=& \left(\begin{array}{cccc}
\frac{\partial \sigma'_1}{\partial \sigma^t_1} &  \multicolumn{2}{c}{\cdots}  & \frac{\partial \sigma'_1}{\partial \sigma^t_k}\\
\vdots & && \vdots\\
\frac{\partial \sigma'_{k-1}}{\partial \sigma^t_1} &   \multicolumn{2}{c}{\cdots} & \frac{\partial \sigma'_{k-1}}{\partial \sigma^t_k}\\
0 & \cdots & 0 & 1\\
\end{array}\right),
\eea
where in the second line we used Eq. (\ref{rec}). We consider Eq. (\ref{lin}) at the critical fixed point $\bsi = \bsi_{\rm c}$, $\beta = \beta_{\rm c}$, solve for  $\partial \sigma'_i/\partial \sigma_j$, and obtain $M$ from Eq. (\ref{M}). The   eigenvalues of $M$ are $\lambda_1 = 1$, and the eigenvalues $\lambda_2, \cdots, \lambda_k$ of the top-left, $(k-1) \times (k-1)$ block of $M$. 

\begin{figure*}
\begin{center}
\includegraphics[scale=1]{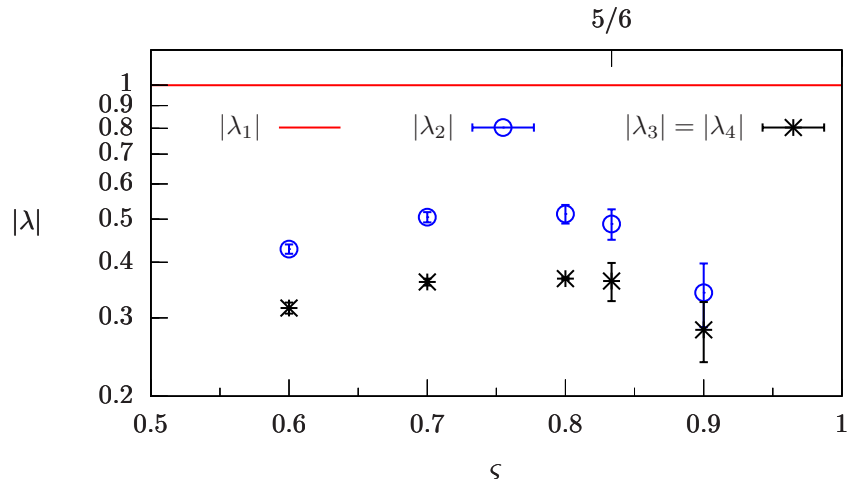}
\caption{\label{lambda}
Norms of the eigenvalues  $\lambda_l$ of the linearized ensemble renormalization-group (ERG) transformation at the critical fixed point as functions of $\varsigma$, for an ERG transformation with $k=4$ hierarchical levels. The plot is on a semi-logarithmic scale, and the value $\varsigma = 5/6$ considered in Figs. \ref{fixed_beta}, \ref{beta_cr} and \ref{spectral} is also marked.}
\end{center}
\end{figure*}

The norms of the eigenvalues resulting from our analysis for $k=4$ are shown in Fig. \ref{lambda} as functions of $\varsigma$. The first eigenvalue $\lambda_1 = 1$ has norm identically equal to one and, because $\lambda_3$ and $\lambda_4$ are complex conjugate to each other, their norms are equal. Importantly, Fig. \ref{lambda} shows that there are no relevant eigenvalues, i.e, for all values of $\varsigma$ considered,
\be
|\lambda_l| \leq 1, \; 1 \leq l \leq 4. 
\ee
By using the known relation between the eigenvalue with the largest norm and $\nu$ \cite{wilson1975renormalization}, we conclude that the ERG approach with $k=4$ hierarchical levels predicts  a value of the critical exponent $
\nu$  given by 
\be\label{nu}
2^{1/\nu} = \lambda_1 = 1,
\ee
which constitutes the main result of our analysis. 

\begin{figure*}
\begin{center}
\includegraphics[scale=1]{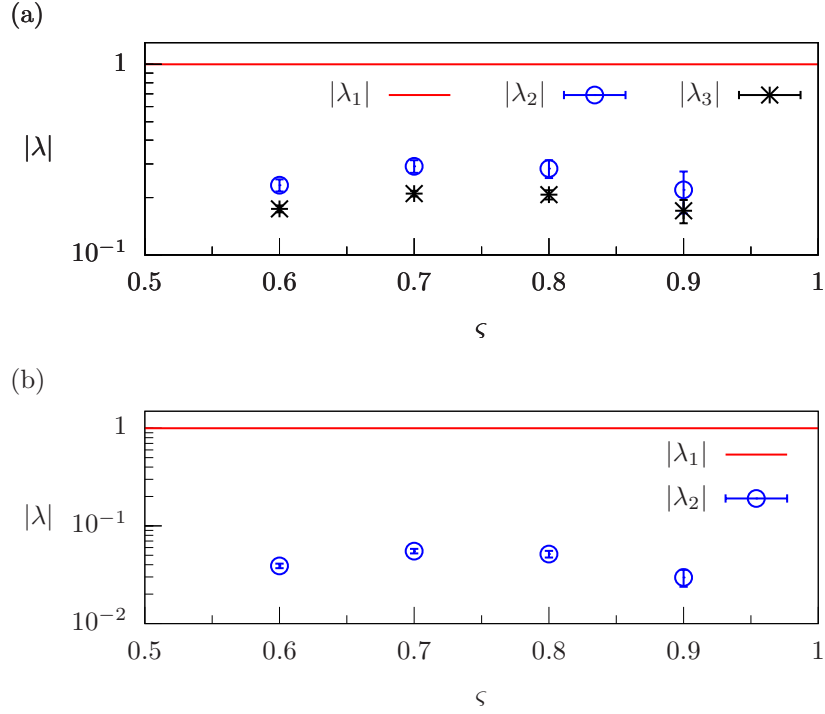}
\caption{\label{lambda_bis}
Norms of the eigenvalues  $\lambda_l$ of the linearized ensemble renormalization-group (ERG) transformation at the critical fixed point as functions of $\varsigma$ on a semi-logarithmic scale, for an ERG transformation with $k=3$ (a) and $k=2$ (b) hierarchical levels.}
\end{center}
\end{figure*}

In addition, in Fig. \ref{lambda_bis} we show the results obtained with the $k=3$ and $k=2$ approximations, thus demonstrating that Eq. (\ref{nu}) holds for  approximations  $k=2,3$ and $4$.

\begin{figure*}
\begin{center}
\includegraphics[scale=.85]{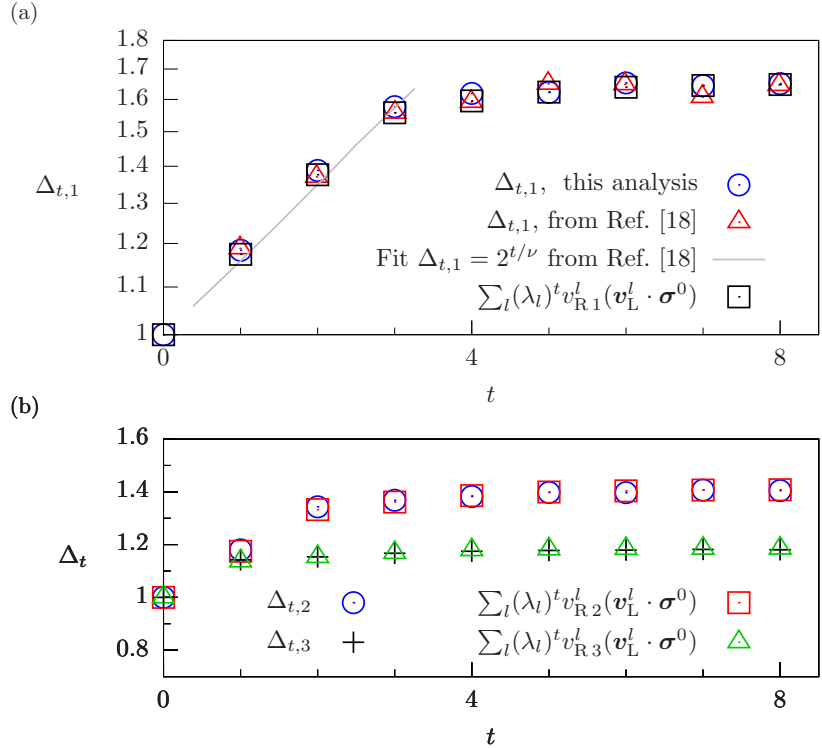}
\caption{\label{spectral}
Divergence between two flows of the ensemble  renormalization-group (ERG) transformation with $k=4$ hierarchical levels at fixed temperatures $T_1 = 0.47$ and $T_2 = 0.57$, with    $\varsigma = 5/6$. 
(a) First component of the divergence, $\Delta_{t,1}$, as a function of the renormalization-group steps $t$, in semi-logarithmic scale. We show $\Delta_{t,1}$ obtained from the present analysis (blue circles), from \cite{angelini2013ensemble} (red triangles), and the fit $\Delta_{t,1} = 2^{t/\nu}$  performed in \cite{angelini2013ensemble} for $1 \leq t \leq 3$  (gray line). Black squares denote the prediction  $\Delta_{t,1}=\sum_l (\lambda_l)^t \,{v}^l_{{\rm R} \, 1}({\bm v}^l_{\rm L} \cdot {\bm \sigma}_0)$ obtained from the linearization of the ERG transformation at the critical fixed point. 
(b)  Second and third component of the divergence, $\Delta_{t,2}$ and $\Delta_{2,3}$ (blue circles and black crosses, respectively). Red squares and green triangles denote the second and third component of the prediction  $\bD_t = \sum_l (\lambda_l)^t \, {\bm v}^l_{{\rm R} }({\bm v}^l_{\rm L} \cdot {\bm \sigma}_0)$ obtained from the linearization of the ERG transformation. All results in (b) are obtained from the present analysis.}
\end{center}
\end{figure*}

\vspace{2mm}

The value of $\nu$ claimed in \cite{angelini2013ensemble} differs from (\ref{nu}), and the reasons for this discrepancy will be discussed in what follows. Rather than linearizing the RG transformation at the critical fixed point, in \cite{angelini2013ensemble} the exponent $\nu$ has been obtained as follows: Given two RG flows $\bsi^{t \,1}$, $\bsi^{t \,2}$ at fixed inverse temperatures $\beta_1$, $\beta_2$, and 
\be
\bD_t \equiv \frac{\beta_1 \bsi^{t \,1} - \beta_2 \bsi^{t \, 2}}{\beta_1 - \beta_2},
\ee
 $\nu$ has  been determined by fitting, for $1 \leq t \leq 3$, the first component of $\bD_t$, i.e., $\Delta_{t,1}$,   with the following formula
\be\label{wil}
\Delta_{t, 1} =  2^{t/\nu},
\ee
see Fig. \ref{spectral}a. However, we observe that the exponential dependence (\ref{wil})  of $\Delta_t$  holds for large $t$ only, not in the region $t \leq 3$ in which it is has been used in \cite{angelini2013ensemble}. This can be demonstrated by observing that, for $\beta \bsi^{t-1} \approx \beta_{\rm c} \bsi_{\rm c}$, we have
\begin{widetext}
\bea\label{eq1}\nn
\beta \sigma^{t}_i - \beta_{\rm c} \sigma_{{\rm c} \, i} &= &F_i[\beta_{\rm c} \bsi_{{\rm c} } + (\beta \bsi^{t-1 } - \beta_{\rm c} \bsi_{{\rm c} })]  -  \beta_{\rm c} \sigma_{{\rm c} \, i}\\ \nn 
& \approx & F_i[\beta_{\rm c} \bsi_{{\rm c} }] +\sum_j \left. \frac{\partial F_i[\beta \bsi]}{\partial (\beta \sigma_j)}\right|_{\beta \bsi = \beta_{\rm c} \bsi_{\rm c}}(\beta \sigma^{t-1 }_j - \beta_{\rm c} \sigma_{{\rm c} \, j})  -  \beta_{\rm c} \sigma_{{\rm c} \, i}\\
&  = &  
\sum_j M_{ij}   (\beta \sigma^{t-1 }_j - \beta_{\rm c} \sigma_{{\rm c} \, j}),
\eea
\end{widetext}
where in the first line we wrote the RG equations in the form $\beta \bsi^{t} = {\bm F}[\beta \bsi^{t-1}]$,  in the second line we expanded  $\bm F$, and in the last line we used the fixed-point condition $ {\bm F}[\beta_{\rm c} \bsi_{\rm c}] = \beta_{\rm c} \bsi_{\rm c}$ and rewrote the derivative of $\bm F$ in terms of $M$ by using Eq. (\ref{M}). By considering Eq. (\ref{eq1}) for $\beta = \beta_1$ and  $\beta = \beta_2$, and  subtracting  side by side, we obtain
\bea\label{eq2}\nn
\bD_{t} & = & \frac{M \cdot( \beta_1 \bsi^{t-1 \; 1}  - \beta_2 \bsi^{t-1 \; 2})}{\beta_1  - \beta_2} \\ \nn
& = & \cdots \\ \nn
& = & \frac{M^{t} \cdot( \beta_1 \bsi^{0 \, 1}  - \beta_2 \bsi^{0 \, 2})}{\beta_1  - \beta_2}\\
& = & \sum_l (\lambda_l)^t \, {\bm v}_{\rm R}^l ({\bm v}_{\rm L}^l \cdot  \bsi^0),
\eea
where in the second and third  line we used recursively Eq. (\ref{eq1})  $t-1$ times. Finally, in the last line we wrote $M$ in terms of its spectral decomposition  
\be
M_{ij} =\sum_l  \lambda_l  v_{{\rm R} \, i}^l v_{{\rm L}\, j}^l,
\ee 
where ${\bm v}_{\rm L}$, ${\bm v}_{\rm R}$ are the left and right eigenvectors of $M$, respectively, and we used the initial condition $\bsi^{0 \, 1} = \bsi^{0 \, 2} \equiv \bsi^0$, where $\bsi^0$ is given by Eq. (\ref{ic}). Equation (\ref{eq2}) shows that the exponential dependence (\ref{wil}) holds only for $t \gg 1$, i.e., in the asymptotic regime  where only the  eigenvalue with the largest norm contributes to $\Delta_{t, 1}$. In addition, Eq. (\ref{eq2})  demonstrates that the exponential form (\ref{wil})  does not hold for small $t$, where all eigenvalues contribute to $\Delta_{t,1}$: it follows that the prediction  for $\nu$ made in \cite{angelini2013ensemble}, which makes use of Eq. (\ref{wil}) for $t \leq3$, is incorrect. 

In Fig. \ref{spectral}a, we illustrate a consistency check of our results, by showing  that the data for  $\bD_{t,1}$ obtained from both our analysis and \cite{angelini2013ensemble} agree with the spectral decomposition (\ref{eq2}).  In particular, it is clear from Fig. \ref{spectral}a that the  increase in $\Delta_{t,1}$ vs. $t$ for $1 \leq t \leq 3$, which in \cite{angelini2013ensemble}  has been interpreted as an exponential increase related to $\nu$ according to Eq. (\ref{wil}),  is actually due to the terms with $2 \leq l \leq 4$ in Eq. (\ref{eq2}), which contain irrelevant eigenvalues. In this regard, it is possible that the apparent maximum in $2^{1/\nu}$ vs. $\varsigma$ at $\varsigma \sim 2/3$ claimed in \cite{angelini2013ensemble} is related to the maximum of the norms of irrelevant eigenvalues $\lambda_2$, $\lambda_3$ and $\lambda_4$, which is  visible in Fig. \ref{lambda}. Finally, Fig. \ref{spectral}b shows that the data for the second and third component of $\bD_t$ obtained from our analysis  also agree with Eq. (\ref{eq2}), thus further validating the spectral decomposition. 

We will now discuss how the correct value of $\nu$ can be recovered by using the fitting method used in \cite{angelini2013ensemble}. We fit $\Delta_{t,1}$---from either our analysis or \cite{angelini2013ensemble}---with Eq. (\ref{eq2}) in  the region $t \gg 1$ where the exponential dependence (\ref{wil}) holds: because $\Delta_{t,1}$ plateaus out for large $t$, by doing so we  recover the  result   obtained with the matrix diagonalization above, i.e., $\lambda_1 = 1 = 2^{1/\nu}$,  which is the correct prediction for the critical exponent $\nu$ resulting from the ERG approach. 

\section{Discussion}

In this analysis, we focused on real-space renormalization-group (RG) methods for the hierarchical Edwards-Anderson model (HEA) \cite{franz2009overlap}---a spin-glass model with long-range interactions built on a hierarchical lattice, where the decay of the interaction strength with respect to the hierarchical distance is set by a parameter, $\varsigma$, which is reminiscent of the space dimension in a short-range system on a hypercubic lattice. 

In a previous study, an RG method based on a sample-by-sample spin decimation has been proposed  \cite{castellana2011real}. This sample-by-sample method yields the correct value of the critical exponent $\nu$  related to the divergence of the correlation length in the region $1/2 < \varsigma < 2/3$ which corresponds to a spatial dimension larger than the upper critical dimension \cite{franz2009overlap,castellana2011renormalization,castellana2015hierarchical}. On the other hand, the prediction  for $\nu$ of the sample-by-sample method disagrees with numerical estimates below the upper critical dimension, i.e., for $\varsigma > 2/3$  \cite{franz2009overlap,angelini2013ensemble}: in particular, the maximum of $2^{1/\nu}$ vs. $\varsigma$ at $\varsigma \sim 2/3$ indicated by numerical simulations is not reproduced by the sample-by-sample method. Further studies proposed a different decimation procedure  in  which, unlike the sample-by-sample method, the decimation involves averages over disorder samples, and claimed that this ensemble RG (ERG) method yields an estimate of $\nu$ in agreement with its mean-field value above the upper critical dimension, and that it reproduces  the maximum of $2^{1/\nu}$ vs. $\varsigma$ at $\varsigma \sim 2/3$ \cite{angelini2013ensemble}. 

In this study, we  analyzed the ERG procedure, in an effort to understand the prediction for $\nu$ claimed in \cite{angelini2013ensemble} and compare it with that of the sample-by-sample method. By  diagonalizing the linearized ERG transformation at the critical fixed point, we demonstrate  that the ERG method with $k=2,3$ and $4$ hierarchical levels yields a marginally stable critical fixed point with no relevant eigenvalues, and that it predicts a value of $\nu$  given by the the simple relation $2^{1/\nu} =1$, which differs from the prediction for $\nu$ claimed in  \cite{angelini2013ensemble}. The cause of this discrepancy is that,  in \cite{angelini2013ensemble},  the exponent $\nu$ has been determined by using the relation (\ref{wil}) in a regime $t \leq 3$, in which such relation does not hold: if the exponential dependence (\ref{wil}) is used in the asymptotic regime $t \gg1 $ where it is valid, then the correct result $2^{1/\nu} =1$ is recovered.

In sum, our analysis demonstrates that the ERG transformation proposed in \cite{angelini2013ensemble} yields, for $k=2,3$ and $4$ hierarchical levels, a prediction for the critical exponent $\nu$ given by $2^{1/\nu} =1$, which disagrees with both the  mean-field value of $\nu$  above the upper critical dimension \cite{franz2009overlap}, and with numerical estimates of $\nu$ below the upper critical dimension \cite{franz2009overlap,angelini2013ensemble}, thus showing that such ERG transformation should be reconsidered. 

In this regard, we observe that a variant of the ERG method has been recently applied to the ferromagnetic version of the HEA with a random magnetic field, where it has been claimed to yield accurate estimates of $\nu$ \cite{decelle2014ensemble}. Specifically, this modification of the ERG method differs from the one of \cite{angelini2013ensemble}. First, when two $k-1$-level systems are recombined so as to form a $k$-level system, the coupling at the $k$th  level is no longer chosen to be equal to the $k$th-level coupling of the original model, cf. Eq. (\ref{rec}), because both the $k$- and $k-1$-level models are assumed to have the same ferromagnetic coupling $J_{ij}=J$ across all hierarchical levels. Second, the exponent $\nu$ is no longer estimated by means of the fitting procedure (\ref{wil}) that lead to an incorrect estimate of $\nu$ in \cite{angelini2013ensemble}, but  by means of a linearization of the RG transformation analog to the one used in our analysis, cf. Eq. (\ref{M}). Finally, it would be interesting to apply the sample-by-sample method to this random-field version of the ferromagnetic hierarchical model, so as to study and compare its predictions for universal quantities with those of the ERG method. 

\vspace{2mm} 
Overall, our analysis indicates that the problem of finding a  real-space RG transformation for a hierarchical spin glass remains an open problem. In particular, we hope that our analysis will provide some general guidance to build one such transformation that yields the correct, quantitative estimates for universal quantities below the upper critical dimension. 

\acknowledgments

We thank A. Barra, M. A. Moore, and F. Zamponi for valuable conversations. The author would like to express his appreciation to I. A. Campbell for many deep and insightful discussions and conversations. This work was granted access to the HPC resources of MesoPSL financed by the Region \^{I}le de France and the project Equip@Meso (reference ANR-10-EQPX-29-01) of the programme Investissements d'Avenir supervised by the Agence Nationale pour la Recherche.

\end{document}